\documentclass[conference,10pt]{IEEEtran}
\usepackage{epsfig,epsf,rotating,setspace,latexsym,amsmath,amssymb,amsfonts,bm,theorem,subfigure,epstopdf,cite,authblk,bbm,nonfloat,comment}
\usepackage{algorithm}
\usepackage[noend]{algpseudocode}
\usepackage{color}
\usepackage{mathtools}
\usepackage{soul}

\usepackage[
top=0.75in,
bottom=1.1in,
left=0.64in,
right=0.65in]{geometry}

\algrenewcommand\algorithmicforall{\textbf{foreach}}
\algrenewcommand\algorithmicindent{.8em}

\newtheorem{lemma}{Lemma}

\newtheorem{remark}{Remark}

\newenvironment{Proof}[1]{\medskip\par\noindent{\bf Proof:\,}\,#1}{{\mbox{\,$\blacksquare$}\par}}

\IEEEoverridecommandlockouts
\allowdisplaybreaks

\begin{document}

\title{Age of Gossip With Cellular Drone Mobility} 

\author{Arunabh Srivastava \qquad Sennur Ulukus\\
        \normalsize Department of Electrical and Computer Engineering\\
        \normalsize University of Maryland, College Park, MD 20742\\
        \normalsize  \emph{arunabh@umd.edu} \qquad \emph{ulukus@umd.edu}}

\maketitle

\begin{abstract}
     We consider a cellular network containing $n$ nodes where nodes within a cell gossip with each other in a fully-connected fashion and a source shares updates with these nodes via a mobile drone. The drone receives source updates and shares them with nodes in the cell where it currently resides. The drone moves between cells according to an underlying continuous-time Markov chain (CTMC). We evaluate the impact of the number of cells $f(n)$, drone speed $\lambda_m(n)$ and drone dissemination rate $\lambda_d(n)$ on the information freshness of nodes in the network. We use the version age of information metric to quantify information freshness. We observe that the expected duration between two drone-to-cell service times depends on the stationary distribution of the underlying CTMC and $\lambda_d(n)$, but not on $\lambda_m(n)$. However, the version age instability makes high probability analysis for a general underlying CTMC difficult. Therefore, we focus on the fully-connected drone mobility model. Under this model, we uncover a dual-bottleneck, by leveraging stochastic equivalence between drone mobility and drone dissemination speed: the version age is constrained by the slower of these two processes. If $\lambda_d(n) \gg \lambda_m(n)$, then the version age scaling of nodes is dominated by the inverse of $\lambda_m(n)$ and is independent of $\lambda_d(n)$. If $\lambda_m(n) \gg \lambda_d(n)$, then the version age scaling of nodes is dominated by the inverse of $\lambda_d(n)$ and is independent of $\lambda_m(n)$.
\end{abstract}

\section{Introduction}\label{sec: introduction}

\begin{figure*}[t]
    \centering
    \includegraphics[width=\linewidth]{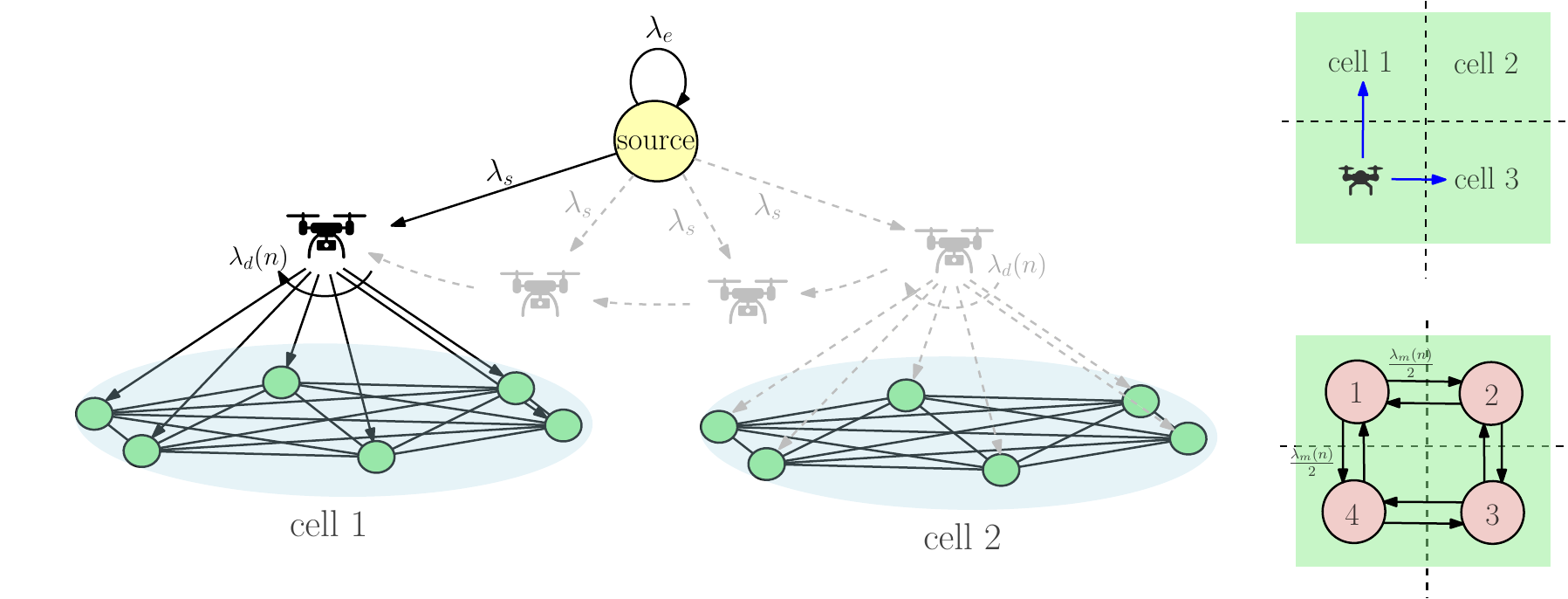}
    \caption{A gossiping network with cellular drone mobility. On the left, a source generates updates and shares them with a mobile drone. The drone moves between cells and disseminates information to all the nodes in a cell as a rate $\lambda_d(n)$ Poisson process. The nodes in each cell gossip as a fully-connected network, but do not gossip with nodes in a different cell. On the right, we see how the drone moves between cells. The top right figure shows which cell the drone can move to from cell $4$ based on the underlying CTMC shown on the bottom right. The holding time of each state of the CTMC has rate $\lambda_m(n)$.}
    \label{fig: system_model_mobility_cellular}
\end{figure*}

Modern wireless networks support a large number of connected devices, and uncrewed aerial vehicles (UAVs) have emerged as an important technology in next generation wireless systems. UAV use for information dissemination in internet of things (IoT) networks has gained significant interest, with applications including environmental IoT sensor networks and disaster response networks. In these applications, drones move between different geographical locations and disseminate or collect information from nodes which can communicate with other nodes in close proximity, but are unable to otherwise communicate with the world. Moreover, these networks operate under time-sensitive constraints, which necessitate the analysis of information freshness.

However, it is not an easy task to maintain information freshness for all nodes in the network with a bandlimited drone which visits clusters of nodes intermittently. Moreover, use of a centralized controller is not possible in such cases. This motivates the inclusion of gossip algorithms \cite{chettri2019comprehensive, swamy2020empirical}, where nodes are able to communicate with close neighbors in their cluster. Gossiping allows every node to receive fresh updates even when the drone is far away and unable to service the cluster. This enables scalable and delay-sensitive information dissemination that ensures freshness despite the nodes receiving few direct updates from the drone.

In such delay-sensitive cellular networks, accurate real-time decision making is dependent on information freshness. Reliance on stale information can cause critical errors and inefficiencies in time-critical applications such as disaster response networks. The need to quantify information freshness has led to the creation of many metrics. The age of information (AoI) metric has been widely adopted as a measure of freshness of information \cite{kaul2012real, sun2019age, yatesJSACsurvey}. Many other metrics have also been proposed, based on important performance indicators. These include the age of incorrect information (AoII) \cite{maatouk20AOII}, the age of synchronization (AoS) \cite{zhong18AoSync}, the binary freshness metric (BFM) \cite{cho3BinaryFreshness}, and the version age of information (version age) \cite{yates21gossip, Abolhassani21version, melih2020infocom}. Significant research has focused on active trajectory optimization to minimize AoI. For instance, \cite{tripathi2019age} investigates trajectory design for information dissemination under random walk mobility, while \cite{wu21uav} and \cite{eslam23uav} utilize reinforcement learning to optimize multi-UAV relaying and energy efficiency. Unlike these works which focus on path planning, we analyze the impact of inherent stochastic mobility models.

In this work, we use the version age of information to quantify information freshness. The version age of information of a node in a gossiping network is defined as the difference between the source version and the node version. This model was first analyzed using the spatial mean field regime method in \cite{chaintreau2009age}, and later using the stochastic hybrid system (SHS) framework in \cite{yates21gossip}. Following works \cite{buyukates22ClusterGossip, kaswan22jamming, kaswan22timestomp, mitra_Infocom23, srivastava2024mobilityagebasedgossipnetworks, maranzatto24} have focused on using the SHS framework and the resulting recursive equations to analyze gossiping networks with diverse properties, including structured topologies, mobility, and adversarial settings. A comprehensive review of these works is provided in \cite{kaswan2025versionagesurvey}. Additionally, \cite{maranzatto2025agegossipconnectiveproperties} characterized the distribution of the version age of information using first passage percolation.

In this work, we consider cellular gossiping networks consisting of $n$ nodes and $f(n)$ cells with equal number of nodes. Information is shared with the nodes by a source via a mobile drone. The nodes in each cell are gossiping in a fully-connected fashion and do not communicate with nodes outside the cell. The mobile drone moves between cells according to an underlying continuous-time Markov chain (CTMC) with rate $\lambda_m(n)$, and disseminates information according to a Poisson process with rate $\lambda_d(n)$. We aim to characterize the average version age of each node as $f(n)$, $\lambda_m(n)$ and $\lambda_d(n)$ vary in $0$ to $\infty$. In Section~\ref{sec: general results}, we first show that the drone's version age is stochastically equivalent to a constant at any time. Next, we show that the expected time between two updates sent by the drone to a cell is dependent on the stationary distribution of the drone's underlying CTMC and $\lambda_d(n)$, but is independent of $\lambda_m(n)$. We also observe that the variance for drone mobility CTMCs prevents us from obtaining high probability results. We then focus our attention to the case of fully-connected drone mobility in Section~\ref{sec: fully-connected drone mobility}, and uncover a dual-bottleneck. If $\lambda_m(n) \gg \lambda_d(n)$, i.e., the network is bandwidth constrained, then the average version age of each node is inversely proportional to $\lambda_d(n)$ and is independent of $\lambda_m(n)$. On the other hand, if $\lambda_d(n) \gg \lambda_m(n)$, in which case the network is mobility constrained, the average version age of each node is inversely proportional to $\lambda_m(n)$ and independent of $\lambda_d(n)$. In both cases, we find that the average version age is directly proportional to $f(n)$.

\section{System Model and the Version Age Metric}\label{sec: system model}
We consider a gossip network where a source generates or observes updates as a rate $\lambda_e$ Poisson process and shares them with a drone as a rate $\lambda_s$ Poisson process. The set of nodes in the gossip network is defined as $\mathcal{N}=\{1,2,\ldots,n\}$, where $n$ is the number of nodes in the gossip network. The entire gossip network is divided into $f(n)$ cells numbered as $1,2, \ldots, f(n)$ in some order without loss of generality. Each cell contains an equal number of nodes. Thus, each cell contains $\frac{n}{f(n)}$ nodes. Moreover, all nodes in each cell are connected in a fully-connected network, i.e., every node in the cell can gossip with every other node in the cell. Gossip between these nodes takes place as a Poisson process. The combined rate of the individual Poisson processes for each node is $\lambda$, and this rate is divided equally among all the neighbors. There is no communication between nodes in different cells. The drone moves between cells and disseminates information to nodes in the cell where it is present. The drone sends updates to the nodes as a combined rate $\lambda_d(n)$ Poisson process, which is equally divided among all the nodes in the cell. We define a continuous-time Markov chain (CTMC) which characterizes the movement of the drone between cells. In this CTMC, the drone moves out of a state with rate $\lambda_m(n)$ irrespective of the cell the drone is currently present in. The next cell which the drone visits is determined by the underlying discrete-time Markov chain. This model is described in Fig.~\ref{fig: system_model_mobility_cellular}. We say that the drone has fully-connected mobility if the drone can move from any cell to any other cell as a rate $\frac{\lambda_m(n)}{f(n)-1}$ Poisson process.

We use the version age of information metric to quantify the freshness of information for the drone and the nodes in the network. We define $N_0(t)$ to be the Poisson counting process (with rate $\lambda_e$), associated with the version update process at the source node. Let $N_0([t_1,t_2])$ be the number of arrivals of the Poisson process between times $t_1$ and $t_2$. Then, $N_0([t_1,t_2])=N_0(t_2)-N_0(t_1)$. Since $N_0(0)=0$, $N_0([0,T])=N_0(T)$, and we use $N_0([0,T])$ and $N_0(T)$ interchangeably. Similarly, we define $N_i(t)$ to be the counting process associated with the version updates at node $i$ in the gossip network. Then, the version age of node $i$ is defined as $X_i(t) = N_0(t) - N_i(t)$. $N_0(t)$ follows a Poisson distribution, but in general, $N_i(t)$ does not, for the gossiping nodes or the drone.

Gossip between nodes within a cell follows the push-based gossiping protocol. Under the push-based protocol, any node sends (pushes) updates to its immediate neighbors at random. If any node receives an update from its neighbor, the node accepts the update if the incoming update is of a better version than the node has. Otherwise, it rejects the packet. Moreover, if the source sends an update to a node, the node's version age drops to $0$. Finally, if the source generates a new version of the update, then every node's version age increases by $1$.

We assume that all Poisson processes are independent of each other. We use the standard asymptotic notation for deterministic scaling, i.e., $O, \Omega, \Theta, o, \omega$. For random variables, we denote stochastic boundedness by $O_p(\cdot)$ and stochastic equivalence by $\Theta_p(\cdot)$.

\section{General Drone Mobility}\label{sec: general results}
In this section, we calculate the version age of each node in the gossiping network. Due to the symmetry of the network, each node within a cell has the same version age experience.

In order to calculate the version age of a node in the gossip network, we calculate how long it takes for an update from the source to reach a node in the gossip network. We divide the travel time $T$ of an update from the source to a node in a cell into three parts. We first calculate $T_{sd}$, the time it takes for an update to reach from the source to the drone. Next, we find the order of the time $T_{dc}$ it takes for an update to reach from the drone to the cell. Finally, we calculate the time $T_{cn}$ for the update to reach every node in the cell, once any node in the cell has the update.

To start, we show that the version age of the drone is $\Theta_p(1)$. This also means that $T_{sd} = \Theta_p(1)$.

\begin{lemma}\label{lemma: 2}
    The version age of the drone is $\Theta_p(1)$.
\end{lemma}

\begin{Proof}
    The inter-arrival time between two source to drone updates is an exponential distribution with rate $\lambda_s$, and the version age of the drone is the number of times the source updates itself since the last update sent by the source to the drone, which has a Poisson distribution. Due to the memoryless property of the exponential distribution, the distribution of the time since the drone last received an update is also an exponential random variable with the same rate. The distribution of the version age of the drone then becomes $\text{Poi}_{\lambda_e}(\text{Exp}(\lambda_s))$. Then, we can find the probability of the version age of the drone $X_d(t)$ at time $t$ being $k$ as follows,
    \begin{align}
        \mathbb{P}[X_d(t)=k] =& \int_0^\infty \frac{(\lambda_e a)^k e^{-\lambda_e a}}{k!} \lambda_s e^{-\lambda_s a} da\\
        =& \frac{\lambda_s}{k!} \int_0^\infty (\lambda_e a)^k e^{-(\lambda_s+\lambda_e)a} da\\
        =& \frac{\lambda_s}{k!}\bigg(\frac{\lambda_e}{\lambda_e+\lambda_s}\bigg)^k \frac{1}{\lambda_e+\lambda_s}\int_0^\infty x^k e^{-x} dx\\
        =& \frac{\lambda_s}{\lambda_e+\lambda_s}\bigg(\frac{\lambda_e}{\lambda_e+\lambda_s}\bigg)^k.
    \end{align}
    Thus, the version age of the drone is distributed as a Geometric distribution with success parameter $\frac{\lambda_s}{\lambda_e+\lambda_s}$. Since $X_d(t)$ is always an integer, and its probability distribution has an exponential tail, it is easy to see that $X_d(t) = \Theta_p(1)$.
\end{Proof}

 Next, we find how long it takes for an update to reach a cell (i.e., any node in the cell is updated by the drone). We wish to find the inter-renewal times of the drone to cell update process. Thus, we first calculate the expectation of the inter-renewal time. We do this in the following lemma.

\begin{lemma}\label{lemma: 3}
    Let the inter-renewal time of the drone to cell update process be $\tau$. Without loss of generality, let the cell be numbered $1$. Then, $\mathbb{E}[\tau] = \frac{1}{\pi_1\lambda_d(n)}$, where $\pi_1$ is the stationary probability of the CTMC associated with cell $1$. Further, $\tau = O_p\big(\frac{1}{\pi_1\lambda_d(n)}\big)$.
\end{lemma}

\begin{Proof}
    Since the process is composed of a mixture of exponential random variables, we use the phase-type (PH) distribution for analysis in this proof. The PH distribution is defined using the starting distribution $\boldsymbol{\alpha}$ and the sub-generator matrix $\boldsymbol{M}$.

    We first see that the inter-renewal time starts right after the drone sends an update to a node in cell $1$. The remaining time the drone spends in cell $1$ is then distributed as an exponential random variable with rate $\lambda_m(n)$, which is the same as the distribution of the time spent by the drone in cell $1$. Moreover, let the absorbing state of the PH distribution be $f(n)+1$. Thus, states $1,2,\ldots, f(n)$ represent states of the CTMC, and $f(n)+1$ represents the state, which if reached, means that the drone has sent an update to cell $1$. Thus, it is clear that a transition to state $f(n)+1$ can be made only from state $1$. Using this information, we can find that $\boldsymbol{M} = \boldsymbol{Q} - \text{diag}(\lambda_d(n),0,\ldots,0)$, where $\boldsymbol{Q}$ is the generator matrix of the CTMC associated with the drone mobility process. Further, $\boldsymbol{\alpha} = \boldsymbol{e}_1$, since we start in state $1$, where $\boldsymbol{e}_1$ is the first standard basis vector. Next, we find the expected value of $\tau$ \cite{bladt2005review}, with defining $\boldsymbol{1} = [1,1,\ldots,1]^T$, as follows
    \begin{align}
        \mathbb{E}[\tau] =& -\boldsymbol{\alpha} \boldsymbol{M}^{-1} \boldsymbol{1}.
    \end{align}
    Calculating $\boldsymbol{M}^{-1}$ directly would not be possible in general. However, we can use a slightly different formulation to calculate the expectation easily. To this end, define $\boldsymbol{x}$ such that
    \begin{align}
        \boldsymbol{M}\boldsymbol{x} = -\boldsymbol{1}.
    \end{align}
    Then, $\mathbb{E}[\tau] = x_1$, the first entry of $\boldsymbol{x}$. Since $\boldsymbol{\pi}$ is the stationary distribution of the CTMC associated with the drone mobility model, we have that $\boldsymbol{\pi}\boldsymbol{Q} = \boldsymbol{0}$. Thus, pre-multiplying by $\boldsymbol{\pi}$,
    \begin{align}
        \boldsymbol{\pi}\boldsymbol{M}\boldsymbol{x} =& \boldsymbol{\pi}(\boldsymbol{Q}-\lambda_d(n)\boldsymbol{E}_{11})\boldsymbol{x}\\
        =&  \boldsymbol{\pi}\boldsymbol{Q}\boldsymbol{x} - \boldsymbol{\pi}\lambda_d(n)\boldsymbol{E}_{11}\boldsymbol{x}\\
        =& - \boldsymbol{\pi}\lambda_d(n)\boldsymbol{E}_{11}\boldsymbol{x}, \label{eq:15}
    \end{align}
    where $\boldsymbol{E}_{11}$ is $1$ in its first diagonal entry, and $0$ in all other entries and is the same shape as $\boldsymbol{Q}$. The left hand side is simply multiplying $\boldsymbol{\pi}$ with $-\boldsymbol{1}$, yielding $-1$. Thus, \eqref{eq:15} simplifies to
    \begin{align}
        \boldsymbol{\pi}\lambda_d(n)\boldsymbol{E}_{11}\boldsymbol{x} = 1. \label{eq:16}
    \end{align}
    Now, since $\boldsymbol{\pi}\boldsymbol{E}_{11} = [\pi_1,0,\ldots,0]$, \eqref{eq:16} simplifies to yield
    \begin{align}
        x_1 = \frac{1}{\pi_1\lambda_d(n)}.
    \end{align}
    Further, a direct application of Markov's inequality yields that $\tau = O_p\big(\frac{1}{\pi_1\lambda_d(n)}\big)$, proving the lemma.
\end{Proof}

This result shows that the speed with which the drone moves between cells does not affect the order of the time (in $n$) between two drone to cell updates. This is a consequence of the memoryless property of the Poisson processes.

Further, we wish to analyze the variance of this inter-renewal time. The variance of $\tau$ is given by
\begin{align}\label{eq: variance}
    \text{Var}[\tau] = 2\boldsymbol{\alpha}\boldsymbol{M}^{-2}\boldsymbol{1} - (\boldsymbol{\alpha} \boldsymbol{M}^{-1} \boldsymbol{1})^2.
\end{align}
In comparison to finding the expectation, finding the variance of $\tau$ is significantly more challenging. Even the order of the variance varies significantly depending on the CTMC structure. Thus, a general high probability result encompassing all mobility patterns is not feasible. We therefore analyze fully-connected mobility in the next section.

\section{Fully-Connected Drone Mobility}\label{sec: fully-connected drone mobility}
In this section, we find the exact value of the variance of $\tau$ when the drone has fully-connected mobility. Finding the variance enables us to apply Lemma~\ref{lemma: 1} to $\tau$, yielding a tight $\Theta_p$ result instead of a loose $O_p$ result for the version age of each node. First, we characterize $\boldsymbol{M}$ in this case as,
\begin{align}
    \!\boldsymbol{M} = \begin{bmatrix}
        -\lambda_m(n)-\lambda_d(n) & \frac{\lambda_m(n)}{f(n)-1} & \ldots & \frac{\lambda_m(n)}{f(n)-1} \\
        \frac{\lambda_m(n)}{f(n)-1} & -\lambda_m(n) & \ldots & \frac{\lambda_m(n)}{f(n)-1}\\
        \vdots & \ddots & & \vdots\\
        \frac{\lambda_m(n)}{f(n)-1} & \ldots & & -\lambda_m(n)
    \end{bmatrix}. 
\end{align}
We can now proceed to find the variance using two systems of linear equations, $\boldsymbol{M}\boldsymbol{x} = -\boldsymbol{1}$ and $\boldsymbol{M}\boldsymbol{y} = -\boldsymbol{x}$. This is done because the variance defined in \eqref{eq: variance} contains a term involving $\boldsymbol{M}^{-2}$, and it is challenging to explicitly calculate $\boldsymbol{M}^{-2}$. The first term in the variance in \eqref{eq: variance} is the second moment of $\tau$. Once we obtain $\boldsymbol{y}$, it is easy to see that $\mathbb{E}[\tau^2]=2y_1$. 

To this end, we first show that in both $\boldsymbol{x}$ and $\boldsymbol{y}$, all entries except the first entry are equal to each other. First, we show that $x_2 = x_3 = \ldots =x_{f(n)}$, and the argument for $\boldsymbol{y}$ then follows in a similar way. Let $S = \sum_{i=2}^{f(n)}x_i$. Note that $S$ can be calculated independently of $x_2, \ldots, x_{f(n)}$ by using the equation from the first row of $\boldsymbol{M}$ and the result for $x_1$ obtained in Lemma~\ref{lemma: 3}, resulting in the fact that $S$ can be treated as a constant in the system of linear equations below. 

Then, we can expand the equation $\boldsymbol{M}\boldsymbol{x} = -\boldsymbol{1}$ for each $i \in \{2, \ldots, f(n)\}$, and write the following from the associated equation corresponding to the respective row of $\boldsymbol{M}$,
\begin{align}
    -\lambda_m(n) x_i + \frac{\lambda_m(n)}{f(n)-1}x_1 + \frac{\lambda_m(n)}{f(n)-1}(S-x_i) = -1.    
\end{align}
Rearranging yields,
\begin{align}
    x_i = \frac{1+\frac{\lambda_m(n)}{f(n)-1}(x_1+S)}{\lambda_m(n)+\frac{\lambda_m(n)}{f(n)-1}}.
\end{align}
We see that the right hand side is independent of all $x_i$, $i \in \{2,\ldots,f(n)\}$. Hence, all $x_i$, $i \in \{2,\ldots,f(n)\}$ are equal. The same effect takes place in the calculation of $y_2,\ldots,y_{f(n)}$, yielding the fact that $y_2 = \ldots = y_{f(n)}$. 

Now, we can calculate $y_1$ explicitly after calculating the intermediate $x_2$. We already know from Lemma~\ref{lemma: 3} that $x_1 = \frac{f(n)}{\lambda_d(n)}$, since the stationary distribution of the CTMC under fully-connected mobility is simply $\big[\frac{1}{f(n)},\ldots,\frac{1}{f(n)}\big]$. Then, $x_2$ can be calculated from the second equation obtained from the first row multiplication in $\boldsymbol{M}\boldsymbol{x}=-\boldsymbol{1}$,
\begin{align}
    (-\lambda_m(n)-\lambda_d(n))x_1+\lambda_m(n)x_2 = -1,
\end{align}
which results in $x_2 = (f(n)/\lambda_d(n))+((f(n)-1)/\lambda_m(n)).$

Next, we use the equations derived from the first two rows of $\boldsymbol{My}=-\boldsymbol{x}$ to find $y_1$ and $y_2$ through,
\begin{align}
    (-\lambda_m(n)-\lambda_d(n))y_1+\lambda_m(n)y_2 \!=& -\frac{f(n)}{\lambda_d(n)}\\
    \frac{\lambda_m(n)}{f(n)\!-\!1}y_1\!+\!\Big(\frac{f(n)\!-\!2}{f(n)\!-\!1}\!-\!1\Big)\lambda_m(n)y_2 \!= &\!-\!\Big(\frac{f(n)}{\lambda_d(n)}\!+\!\frac{f(n)\!-\!1}{\lambda_m(n)}\Big).
\end{align}
Solving these equations simultaneously yields the following,
\begin{align}
    y_1 = \frac{(f(n))^2}{(\lambda_d(n))^2}+\frac{(f(n)-1)^2}{\lambda_d(n)\lambda_m(n)}.
\end{align}
Substituting this in the variance formula \eqref{eq: variance} along with the value of the expectation $\mathbb{E}[\tau] = x_1$, and the second moment $\mathbb{E}[\tau^2] = 2y_1$, we obtain,
\begin{align}\label{eq: fully connected variance}
    \text{Var}[\tau] = \frac{(f(n))^2}{(\lambda_d(n))^2}+\frac{2(f(n)-1)^2}{\lambda_d(n)\lambda_m(n)}.
\end{align}

We now discuss the two regimes generated by this calculation. Since the source generates updates during the random time window $T$ when the drone is not sending updates to the cell, the number of version updates during this time is exactly $N_0(T)$. During this window, the cell's version age increases by $N_0(T)$ before the drone delivers fresh updates to the cell. We first identify the window $T$, its statistical properties in each regime and its asymptotic distribution in each regime. 

\textbf{Regime 1}: $\lambda_m(n)=\omega(\lambda_d(n))$. In this regime, the drone moves between cells faster than or at the same rate as it updates them. In this case, we infer from \eqref{eq: fully connected variance} and Lemma~\ref{lemma: 3} that $\text{Var}[\tau] = O((\mathbb{E}[\tau])^2)$. This relation between the variance and squared mean implies that the random time window $T$ is $\tau$, since it represents the bottleneck of the drone successfully updating the cell. We conclude that the time between two successive drone-to-cell updates is $T_{dc} = \Theta_p(\mathbb{E}[\tau]) = \Theta_p(\frac{f(n)}{\lambda_d(n)})$. Structurally, this inter-renewal time is a geometric sum of network excursions. This is because, every time we visit state 1, the drone disseminates information with probability $\frac{\lambda_d(n)}{\lambda_d(n)+\lambda_m(n)} \rightarrow 0$, and otherwise goes out of the cell. This process continues until the drone successfully sends an update to the cell. By the generalized Renyi theorem for geometric sums \cite[Chapter~3]{kalashnikov2013geometric}, as the network scales, $\frac{\tau}{\mathbb{E}[\tau]} \xrightarrow{d} \text{Exp}(1)$.
    
\textbf{Regime 2}: $\lambda_m(n)=O(\lambda_d(n))$. In this regime, the drone updates the cell significantly faster than it moves between cells. Consequently, $\text{Var}[\tau]$ scales as $\omega(1)$ relative to $(\mathbb{E}[\tau])^2$. This indicates that the updates arrive in heavy bursts. In other words, when the drone arrives in the cell, it sends out numerous updates to the nodes in the cell with high probability before leaving. Due to this behavior, the inter-renewal time no longer accurately describes the true bottleneck. Instead, the bottleneck $T$ shifts to the return time of the drone to the cell $T_{ret}$. Once the drone arrives at a cell, the cell's minimum version age mimics that of the drone, and becomes $\Theta_p(1)$ behind compared to the drone, and thus also the source. By modeling the return time of the drone to the cell as hitting times in the fully-connected mobility CTMC, we see that the expected return time is $\Theta_p\big(\frac{f(n)}{\lambda_m(n)}\big)$ and the associated variance is proportional to the square of this expected value. Further, the return time is the sum of an exponential holding time and a hitting time. Due to the symmetry of the drone CTMC, the hitting time has an exponential distribution with rate $\frac{\lambda_m(n)}{f(n)-1}$. This is because the rest of the network can be treated as a single state from which the time to hit the required state is exponentially distributed. Further, since the expected hitting time dominates the holding time as $f(n) \rightarrow \infty$, $\frac{T_{ret}}{\mathbb{E}[T_{ret}]} \xrightarrow{d} \text{Exp}(1)$. 

We now have the following lemma, which helps us find stochastic equivalence results for $N_0(T)$ for both regimes.

\begin{lemma}\label{lemma: 1}
    Let $T$ be a random time window such that $\mathbb{E}[T] = \Theta(g(n))$ and $\frac{T}{\mathbb{E}[T]} \xrightarrow{d} \text{Exp}(1)$. Then, $N_0(T) = \Theta_p(g(n))$.
\end{lemma}

\begin{Proof}
    We use Markov's inequality to prove the upper bound. First, we calculate $\mathbb{E}[N_0(T)] = \mathbb{E}[\mathbb{E}[N_0(T)|T]] = \mathbb{E}[\lambda_e T] = \lambda_e\mathbb{E}[T]$. Next, to show stochastic boundedness, given $\epsilon>0$, we find finite $M>0$ and finite integer $N_1>0$ such that,
    \begin{align}
        \mathbb{P}\bigg(\bigg|\frac{N_0(T)}{g(n)}\bigg|>M\bigg) < \epsilon \quad \forall n \geq N_1.
    \end{align}
    We first choose $N_1$ such that $c_1 g(n) \leq \mathbb{E}[T] \leq c_2 g(n), \forall n \geq N_1$. Using Markov's inequality, we have,
    \begin{align}
        \mathbb{P}\bigg(\bigg|\frac{N_0(T)}{g(n)}\bigg|>M\bigg) \leq& \frac{\mathbb{E}[N_0(T)]}{Mg(n)}
        =& \frac{\lambda_e\mathbb{E}[T]}{Mg(n)}
        \leq& \frac{\lambda_e c_2}{M}.
    \end{align}
    Choosing $M > \frac{\lambda_e c_2}{\epsilon}$ completes the proof of the first part. 
    
    To prove the lower bound, we evaluate the asymptotic distribution of $T$. $N_0(T)$ represents Poisson arrivals during $T$, which is asymptotically exponentially distributed. Therefore, $N_0(T)$ asymptotically follows a geometric distribution, as we proved in Lemma~\ref{lemma: 2}. Further, $\mathbb{E}[N_0(T)] \rightarrow \infty$ as $n \rightarrow \infty$. Hence, we can apply the continuous limit of the geometric distribution. This means that $X_n = \frac{N_0(T)}{\mathbb{E}[N_0(T)]}$ converges in distribution to $\text{Exp}(1)$. Let $F_n(x)$ be the cumulative distribution function (CDF) of $X_n$, and $G(x)$ be the CDF of $\text{Exp}(1)$. 

    Then, by the definition of convergence in distribution, 
    \begin{align}
        \sup_{x \geq 0} | F_n(x) - G(x) | \leq \Delta_n, \quad \text{with } \lim_{n \to \infty} \Delta_n = 0.
    \end{align}
    where $\Delta_n$ is the error term. Thus, for any constant $m > 0$, 
    \begin{align}\label{eq: omega p upper bound}
        \mathbb{P}\big(X_n \!\leq\! m\big) \!&=\! \mathbb{P}\big(N_0(T)\! \leq\! m \mathbb{E}[N_0(T)]\big)\!\leq\! 1\! - \!e^{-m}\! + \!\Delta_n.
    \end{align}
    To prove the $\Omega_p$ bound, let $\epsilon>0$ be given. First, we choose $m>0$ such that $1-e^{-m} < \frac{\epsilon}{2}$. This requires $m < -\ln\left(1-\frac{\epsilon}{2}\right)$. Next, since $\Delta_n \rightarrow 0$, there exists finite positive integer $N_2$ such that $\forall n\geq N_2$, $\Delta_n < \frac{\epsilon}{2}$. Therefore, continuing from \eqref{eq: omega p upper bound}, for all $n \geq N_2$, we have,
    \begin{align}\label{eq: omega p final eqn}
        \mathbb{P}(N_0(T) \leq m \mathbb{E}[N_0(T)]) < \frac{\epsilon}{2}+\frac{\epsilon}{2} = \epsilon.
    \end{align}
    This shows that $N_0(T) = \Omega_p(g(n))$, since we have found the required $m$ and $N_2$ so that \eqref{eq: omega p final eqn} holds. This proves the lower bound, and the lemma.
\end{Proof}

Now that we have proven Lemma~\ref{lemma: 1}, we can determine the version age increase for both regimes during the time $T_{dc}$. In Regime 1, applying the lemma to $T = \tau$ with $g(n) = \frac{f(n)}{\lambda_d(n)}$ yields that the version age increase scales as $\Theta_p\big(\frac{f(n)}{\lambda_d(n)}\big)$. In Regime 2, applying Lemma~\ref{lemma: 1} to $T = T_{ret}$ with $g(n) = \frac{f(n)}{\lambda_m(n)}$ yields that the version age increase by $\Theta_p\big(\frac{f(n)}{\lambda_m(n)}\big)$ during the drone's absence.

Finally, we discuss how fast an update is able to spread to the nodes in the cell once a single node has received the update from the drone by evaluating $T_{cn}$. It was shown in \cite{srivastava2024varyingtopologies} that any packet that arrives in a fully-connected network of size $\frac{n}{f(n)}$ reaches every node in the network in $\Theta\big(\log{\frac{n}{f(n)}}\big)$ time w.h.p. Thus, the version age of any node in the cell is $O_p\big(\log{\frac{n}{f(n)}}\big)$ behind when compared to the minimum version age of the cell. Moreover, while a small subset of nodes receives the updates faster, for all but a vanishing fraction $o\big(\frac{n}{f(n)}\big)$ of nodes, the gossip delay scales as $\Theta_p\big(\log\frac{n}{f(n)}\big)$. Thus, the version age of an arbitrarily selected node in the cell will lag behind the cell's minimum version age by $\Theta_p\big(\log\frac{n}{f(n)}\big)$.

In the first regime, the update time from the drone to each node in the gossip network is $\Theta_p\big(\frac{n}{f(n)}\times \frac{f(n)}{\lambda_d(n)}\big) = \Theta_p\big(\frac{n}{\lambda_d(n)}\big)$. Thus, if $\frac{n}{\lambda_d(n)} = o\big(\log{\frac{n}{f(n)}}\big)$, then the version age of each node will be $\Theta_p\big(\frac{n}{\lambda_d(n)}\big)$, since the drone sends updates to the network faster than the nodes in the cell share the update with each other by gossiping. If this is not the case, then the version age of all but a vanishing fraction of nodes in the network is $\Theta_p\big(\frac{f(n)}{\lambda_d(n)}+\log{\frac{n}{f(n)}}\big)$.

In the second regime, the time it takes for all but a vanishing fraction of nodes to receive an update from the source is $\Theta_p\big(\frac{f(n)}{\lambda_m(n)}+\log{\frac{n}{f(n)}}\big)$ due to the drone return time being the dominant delay, and from Lemma~\ref{lemma: 1}, we can conclude the same for each node's version age. The only exception is when the drone to cell update rate is superlinear and the return time of the drone to the cell is sublogarithmic. In this case, every time the drone visits the cell, the drone sends an update to every node w.h.p., resetting their version age to $\Theta_p(1)$. Then, the drone leaves and returns in sublogarithmic time. In this case, the version age scaling is $\Theta_p\big(\frac{f(n)}{\lambda_m(n)}\big)$.

In summary, we have two regimes:\\
\textbf{Regime 1 ($\lambda_m(n) = \omega(\lambda_d(n))$):} 
    If $\frac{n}{\lambda_d(n)} = O\left(\log{\frac{n}{f(n)}}\right)$, the drone updates individual nodes faster than the gossip mechanism spreads information. In this case, the version age of an arbitrary node is $\Theta_p\left(\frac{n}{\lambda_d(n)}\right)$. Otherwise, the version age scaling for all but a vanishing fraction of nodes is $\Theta_p\left(\frac{f(n)}{\lambda_d(n)} + \log{\frac{n}{f(n)}}\right)$.

\noindent\textbf{Regime 2 ($\lambda_m(n) \!\!= \!\!O(\lambda_d(n))$):} 
    If $\frac{n}{\lambda_d(n)}\!\! = \!\!o(1)$ and $\frac{f(n)}{\lambda_m(n)}\!\! = \!\!o\left(\log{\frac{n}{f(n)}}\right)$, every node is reset to a constant version age w.h.p. during each drone visit, and the version age of an arbitrary node is $\Theta_p\!\!\left(\!\frac{f(n)}{\lambda_m(n)}\!\right)$. Otherwise, the version age scaling for all but a vanishing fraction of nodes is $\Theta_p\!\!\left(\!\frac{f(n)}{\lambda_m(n)} \!+\! \log{\frac{n}{f(n)}}\!\right)$.


\section{Remarks}

\begin{remark}
    If there was no drone in the network, and the source was sending updates to each node with equal rate, then the topology of the entire network would remain constant, and we can directly apply the result found for the fully-connected network in \cite{yates21gossip} here. This yields the result that the long-term average version age of every node in the network scales as $\frac{\lambda_e}{\lambda}f(n)\log{\frac{n}{f(n)}}$.
\end{remark}

\begin{remark}
    A similar model was discussed in \cite{buyukates22ClusterGossip}, where nodes were divided into clusters in a fully-connected fashion. Further, each cluster is serviced by a cluster head, which unlike the drone, sends updates to each node continuously. It was observed that the long-term average version age of a node in such a network was $O(f(n)+\log{\frac{n}{f(n)}})$, following a similar structure to our result. However, in our model, the version age of each node has an explicit dependence on either $\lambda_m(n)$ or $\lambda_d(n)$. This shows that mobility and drone dissemination rate affect the version age in our model. We observe that $\lambda_m(n)$ has no equivalent effector in their model. Moreover, the effects of $\lambda_d(n)$ are significant in our model, but it does not affect the version age in \cite{buyukates22ClusterGossip} unless it is $\Omega(n)$.
\end{remark}

\begin{remark}
    We observe that if $f(n) = 1$, we return to the fully-connected network and the resultant version age of each node is $\Theta_p(\log{n})$. This agrees with the result proved in \cite{yates21gossip}, where the long-term average version age was shown to be exactly $\log{n}$. Further, if $f(n) = n$, then each cell will have only one node, and the benefits of gossip will be lost. In this case, the version age of each node is given as $\Theta_p\big(\max\big(\frac{n}{\lambda_m(n)}, \frac{n}{\lambda_d(n)}\big)\big)$.
\end{remark}

\bibliographystyle{unsrt}
\bibliography{refs}

\end{document}